\documentstyle[preprint,prc,eqsecnum,aps,epsf]{revtex}

\tightenlines

\newcommand{\ffat}[1]{\mbox {\boldmath $#1$}}

\begin{document}
\draft

\title{New Forms of Deuteron Equations and Wave Function Representations}

\author{ I.~Fachruddin$^{\dagger}$\thanks{ Permanent address: Jurusan Fisika, FMIPA, 
Universitas Indonesia, Depok 16424, Indonesia}, Ch.~Elster$^{\ddagger}$,
W.~Gl\"ockle$^{\dagger}$} 
\address{ $^{\dagger}$Institut f\"ur Theoretische Physik II, 
Ruhr-Universit\"at Bochum, D-44780 Bochum, Germany.}
\address{$^{\ddagger}$ Institut f\"ur Kernphysik, Forschungszentrum
J\"ulich, D-52425 J\"ulich, and
Institute of Nuclear and
Particle Physics, Ohio University,
Athens, OH 45701}

\vspace{10mm}

\date{\today}

\maketitle

\begin{abstract}
A recently developed helicity basis for nucleon-nucleon (NN) scattering is applied to the
deuteron bound state.  Here the total spin of the deuteron is treated
in such a helicity representation. For the bound state,
two  sets of two coupled eigenvalue equations are developed, 
where the amplitudes depend on two and one variable, respectively. 
Numerical illustrations based on the realistic Bonn-B NN potential are given. 
In addition, an `operator form' of the deuteron wave function is
presented, and
several momentum dependent spin densities are derived and shown, in which
the angular dependence is given analytically.
\end{abstract}

\vspace{10mm}

\pacs{PACS number(s): 13.75.Cs, 21.10.Hw, 21.45.+v, 27.10.+h} 

\pagebreak
 
\narrowtext

\section{Introduction}

In a recent manuscript \cite{1} we developed a three-dimensional approach in momentum space 
for nucleon-nucleon (NN) scattering. 
The motivation is that for higher energies too many partial waves 
contribute and a direct solution seems more natural and 
economic. As relevant variables momentum vectors appear, specifically their magnitudes and the
angles between them.
The formulation in \cite{1} is based on a helicity representation with respect to 
the total  spin of the two nucleon system. 
This representation  is different from the often used helicity basis 
referring to the individual nucleons \cite{Kubis,alzetta}. 
A further important advantage of a three 
dimensional approach is that a sometimes tedious partial wave expansion of a complex NN force 
is no longer needed. Instead one introduces a helicity representation of the NN force, 
which is perfectly adapted 
to the set of six operators completely describing the most general NN force compatible
with general invariance principles.
Thus, for any NN force given in operator form this scheme 
is applicable. 

The helicity representation developed for NN scattering can also be applied to
the bound NN system.
It may appear unnecessary to extend this  particular formulation to study 
the nonrelativistic deuteron,  which only contains $S$ and $D$ waves. 
However, the standard practice
requires a partial wave representation of the NN force, which we avoid.
It is straightforward to extend the helicity formulation developed for
scattering 
to investigate the deuteron, calculate its binding energy,  
and wave function properties. This is the purpose of the present investigation. In addition we
study the 
various wave function properties in momentum space in a three dimensional fashion. 
Our numerical example is based on the Bonn-B potential \cite{bonnb}. 
A graphical study of similar character has been carried out in configuration space \cite{4} based
on the AV18 nuclear force \cite{av18}. In addition we derive and display 
probability densities of various spin configurations for an overall polarized 
deuteron. These densities are based  on analytical 
expressions, which we think are new.

In Section II we introduce the expansion of the deuteron state into  helicity
basis states as defined in \cite{1}. Then we project the deuteron eigenvalue equation
on these states. Since the deuteron has spin 1 there are three
possible values for the helicity projections, namely $ \Lambda =1,\, 0,\, -1 $.
Symmetry properties  allow to consider only $ \Lambda =1,\, 0 $.
Thus we obtain a set of two coupled equations in two variables, 
the magnitude $|{\bf q}|$ of the relative momentum vector and the angle 
between  ${\bf q}$ and the arbitrarily chosen z-axis. 
This two dimensional form of the deuteron wave function 
is then connected to the standard partial wave representation.
We demonstrate that this set of two coupled 
two dimensional equations can be readily solved and display various wave function 
properties.

In Section III we derive the deuteron wave function in `operator form'. In a
configuration space representation such a form has been given before \cite{ericson}.
For our purpose an `operator form' 
is an ideal starting point, since the spin degrees of freedom appear 
explicitly as spin operators, and thus fit perfectly into our helicity formulation. 
The projection
of the wave function  on the helicity basis leads to deuteron wave function
components with an analytical angular behavior, which is different from the familiar one. The
`radial' part of the wave function satisfies a set of two coupled ($ \Lambda =1,\, 0 $),
one dimensional eigenvalue equations in $ |{\bf q}| $. Based on this more 
analytical insight  the connection to the standard $S$ and $D$ 
waves forms started in Section II can be finalized. 

There are various possibilities that the two nucleons in the deuteron have a
specific  orientation of their spins for an overall polarized deuteron. 
For instance, both nucleons can have their spins up, or
one nucleon can have its spin up and the other down. In Section IV we derive
analytic expressions of the corresponding probabilities and display
the results.
This may have applications for electron scattering on the deuteron.
Finally we summarize in Section V.


\section{Formulation I}

\subsection{Deuteron Wave Function in the Helicity Basis}

Let $ | \Psi _{d}^{M_{d}}\rangle$  represent the deuteron state.
Here $M_{d}$ is the projection of the total angular momentum along a chosen
axis, e.g. the  z-axis. The deuteron state will now be represented in
 the helicity basis 
$ \left| {\bf q};\hat{q}S\Lambda ;t\right\rangle ^{\pi a} $ 
defined in \cite{1}. Here $\bf{q} $ stands for the relative momentum 
of the two nucleons, 
$S$ for the total spin, $\Lambda$ for its projection along $\bf{q}$, and
$t$ for the total isospin. The index $\pi$ denotes the parity of the
state, and $a$ indicates the state being antisymmetric.
This results in
\begin{eqnarray}
\left| \Psi _{d}^{M_{d}}\right\rangle  
& = & \frac{1}{4}\sum ^{1}_{\Lambda =-1}\int d{\bf q}\left| {\bf
q};\hat{q}1\Lambda ;0\right\rangle ^{1a}\left. ^{^{}}\right.
^{1a}\left\langle {\bf q};\hat{q}1\Lambda ;0\right| \left. \Psi
_{d}^{M_{d}}\right\rangle.  \label{eq2.1} 
\end{eqnarray}
Here we took into account that for the deuteron $S=1$, $t=0$ and the
parity is even.
The general form of the helicity eigenstate is given by
\cite{1}
\begin{eqnarray}
\left| {\bf q};\hat{q}S\Lambda ;t\right\rangle ^{\pi a} & = & 
\left( \left| {\bf q}\right\rangle +\eta _{\pi }\left| -{\bf q}\right\rangle 
\right) \left| \hat{q}S\Lambda \right\rangle \left| t\right\rangle , 
\label{label3} 
\end{eqnarray}
where $ \eta _{\pi} $ denotes the parity eigenvalue. Using $ \left| -\hat{q}S\Lambda
\right\rangle =(-)^{S}\left| \hat{q}S-\Lambda \right\rangle $ one verifies
 the following properties:
\begin{eqnarray}
\left| {\bf q};\hat{q}S\Lambda ;t\right\rangle ^{\pi a} & = & \eta _{\pi }(-)^{S}\left( \left| -{\bf q}\right\rangle +\eta _{\pi }\left| {\bf q}\right\rangle \right) \left| -\hat{q}S-\Lambda \right\rangle \left| t\right\rangle \nonumber \\
 & = & \eta _{\pi }(-)^{S}\left| -{\bf q};-\hat{q}S-\Lambda ;t\right\rangle ^{\pi a}.
\end{eqnarray}
With the above relations one obtains for the integral in
Eq.~(\ref{eq2.1})
\begin{eqnarray}
\int d{\bf q}\: | {\bf q};\hat{q}1-1;0\rangle ^{1a} 
 \; ^{1a}\langle {\bf q};\hat{q}1-1;0| \Psi _{d}^{M_{d}}\rangle
 & =  & 
\int d{\bf q} \: | -{\bf q};-\hat{q}11;0\rangle ^{1a} \;
 ^{1a}\langle -{\bf q};-\hat{q}11;0| \Psi _{d}^{M_{d}}\rangle \nonumber \\
 & = & \int d{\bf q} \: | {\bf q};\hat{q}11;0\rangle ^{1a} \;
 ^{1a}\langle {\bf q};\hat{q}11;0 |  \Psi _{d}^{M_{d}}\rangle .
\end{eqnarray}
Hence,  Eq.~(\ref{eq2.1}) simplifies to
\begin{eqnarray}
\left| \Psi _{d}^{M_{d}}\right\rangle  & = & \int d{\bf q}\left\{ \frac{1}{2}\left| {\bf q};\hat{q}11;0\right\rangle ^{1a}\left. ^{^{}}\right. ^{1a}\left\langle {\bf q};\hat{q}11;0\right| \left. \Psi _{d}^{M_{d}}\right\rangle \right. +\left. \frac{1}{4}\left| {\bf q};\hat{q}10;0\right\rangle ^{1a}\left. ^{^{}}\right. ^{1a}\left\langle {\bf q};\hat{q}10;0\right| \left. \Psi _{d}^{M_{d}}\right\rangle \right\} \nonumber \\
 & \equiv  & \int d{\bf q}\left\{ \frac{1}{2}\left| {\bf q};\hat{q}11;0\right\rangle
^{1a}\varphi ^{M_{d}}_{1}({\bf q})+\frac{1}{4}\left| {\bf q};\hat{q}10;0\right\rangle
^{1a}\varphi ^{M_{d}}_{0}({\bf q})\right\} ,
\end{eqnarray}
where we defined
\begin{equation}
\label{label11}
\varphi ^{M_{d}}_{\Lambda }({\bf q})\equiv \left. ^{^{}}\right. ^{1a}\left\langle {\bf
q};\hat{q}1\Lambda ;0\right| \left. \Psi _{d}^{M_{d}}\right\rangle .
\end{equation}

The azimuthal dependency of the amplitude defined in Eq.~(\ref{label11}) can be found as follows. 
The state $\left| {\bf q};\hat{q}S\Lambda \right\rangle$ is obtained by 
rotating the state $| q\hat{z};\hat{z}S\Lambda \rangle$ from the z-axis
into the direction of $\bf q$ as
\begin{equation}
\left| {\bf q};\hat{q}S\Lambda \right\rangle  = R(\hat{q})\left|
q\hat{z};\hat{z}S\Lambda \right\rangle,
\end{equation}
where $ R(\hat{q}) = \exp{-iJ_{z}\phi} \exp{-iJ_{y}\theta }$, 
and $ {\bf J} = {\bf L}+{\bf S} $ is the operator of total angular momentum. It follows that
\begin{eqnarray}
\left. ^{^{}}\right. ^{1a}\left\langle {\bf q};\hat{q}S\Lambda ;t\right| \left. \Psi _{d}^{M_{d}}\right\rangle  & = & \left. ^{^{}}\right. ^{1a}\left\langle q\hat{z};\hat{z}S\Lambda ;t\right| e^{iJ_{y}\theta } e^{iJ_{z}\phi }\left| \Psi _{d}^{M_{d}}\right\rangle \nonumber \\
 & = & e^{iM_{d}\phi }\left. ^{^{}}\right. ^{1a}\left\langle q\hat{z};\hat{z}S\Lambda
;t\right| e^{iJ_{y}\theta }\left| \Psi _{d}^{M_{d}}\right\rangle .
\end{eqnarray}
Thus, we can redefine $ \varphi ^{M_{d}}_{\Lambda }({\bf q}) $ such that the
azimuthal dependency is factored out
\begin{equation}
\label{label29}
\varphi ^{M_{d}}_{\Lambda }({\bf q})\equiv \varphi ^{M_{d}}_{\Lambda }(q,\theta )
e^{iM_{d}\phi } .
\end{equation}
This leads to the final expression of the deuteron state in the helicity basis
\begin{equation}
\label{label2}
\left| \Psi _{d}^{M_{d}}\right\rangle =\int d{\bf q}\left\{ \frac{1}{2}\left|
 {\bf q};\hat{q}11;0\right\rangle ^{1a}\varphi ^{M_{d}}_{1}(q,\theta )+
\frac{1}{4}\left| {\bf q};\hat{q}10;0\right\rangle ^{1a}
\varphi ^{M_{d}}_{0}(q,\theta )\right\} e^{iM_{d}\phi }.
\end{equation}

\noindent
The normalization of the wave function components 
$\varphi ^{M_{d}}_{\Lambda }(q,\theta ) $ can be  calculated as
\begin{eqnarray}
\left\langle \Psi _{d}^{M_{d}}\right| \left. \Psi _{d}^{M_{d}}\right\rangle & = & \int d{\bf q}'\int d{\bf q}\left\{ \frac{1}{2}\left. ^{^{}}\right. ^{1a}\left\langle {\bf q}';\hat{q}'11;0\right| \varphi ^{M_{d}\ast }_{1}(q',\theta ')+\frac{1}{4}\left. ^{^{}}\right. ^{1a}\left\langle {\bf q}';\hat{q}'10;0\right| \varphi ^{M_{d}\ast }_{0}(q',\theta ')\right\} \nonumber \\
 &  & \left\{ \frac{1}{2}\left| {\bf q};\hat{q}11;0\right\rangle ^{1a}\varphi ^{M_{d}}_{1}(q,\theta )+\frac{1}{4}\left| {\bf q};\hat{q}10;0\right\rangle ^{1a}\varphi ^{M_{d}}_{0}(q,\theta )\right\} \nonumber \\
 & = & \int d{\bf q}'\int d{\bf q}\left\{ \frac{1}{4}\left. ^{^{}}\right. ^{1a}\left\langle {\bf q}';\hat{q}'11;0\right| \left. {\bf q};\hat{q}11;0\right\rangle ^{1a}\varphi ^{M_{d}\ast }_{1}(q',\theta ')\varphi ^{M_{d}}_{1}(q,\theta )\right. \nonumber \\
 &  & +\frac{1}{16}\left. ^{^{}}\right. ^{1a}\left\langle {\bf q}';\hat{q}'10;0\right| \left. {\bf q};\hat{q}10;0\right\rangle ^{1a}\varphi ^{M_{d}\ast }_{0}(q',\theta ')\varphi ^{M_{d}}_{0}(q,\theta )\nonumber \\
 &  & +\frac{1}{8}\left. ^{^{}}\right. ^{1a}\left\langle {\bf q}';\hat{q}'11;0\right| \left. {\bf q};\hat{q}10;0\right\rangle ^{1a}\varphi ^{M_{d}\ast }_{1}(q',\theta ')\varphi ^{M_{d}}_{0}(q,\theta )\nonumber \\
 &  & +\left. \frac{1}{8}\left. ^{^{}}\right. ^{1a}\left\langle {\bf q}';\hat{q}'10;0\right| \left. {\bf q};\hat{q}11;0\right\rangle ^{1a}\varphi ^{M_{d}\ast }_{0}(q',\theta ')\varphi ^{M_{d}}_{1}(q,\theta )\right\} \nonumber \\
 & = & 2\pi \int _{0}^{\infty }dq\, q^{2}\int _{-1}^{1}d\cos\theta \, \left\{
\frac{1}{2}\left| \varphi ^{M_{d}}_{1}(q,\theta )\right| ^{2}+\frac{1}{4}\left| \varphi
^{M_{d}}_{0}(q,\theta )\right| ^{2}\right\} \label{label22} .
\end{eqnarray}
Here we used that 
\begin{eqnarray}
\left. ^{^{}}\right. ^{\pi 'a}\left\langle {\bf q}';\hat{q}'S'\Lambda ';t'\right| \left. {\bf q};\hat{q}S\Lambda ;t\right\rangle ^{\pi a} & = & (1-\eta _{\pi }(-)^{S+t})\delta _{t't}\delta _{\eta _{\pi '}\eta _{\pi }}\delta _{S'S} \nonumber\\
&  & \left( \delta ({\bf q}'-{\bf q})\delta _{\Lambda '\Lambda }+\eta
_{\pi }(-)^{S}\delta ({\bf q}'+{\bf q})
\delta _{\Lambda ',-\Lambda }\right). 
\end{eqnarray}

\noindent
From the normalization calculated using Eq.~(\ref{eq2.1}) we
define a deuteron momentum density $ \rho ^{M_{d}}({\bf q}) $ as
\begin{eqnarray}
\rho ^{M_{d}}({\bf q}) & \equiv & 
 \frac{1}{4}\left| \varphi ^{M_{d}}_{1}({\bf q})\right| ^{2}
+\frac{1}{4}\left| \varphi ^{M_{d}}_{0}({\bf q})\right| ^{2} 
+\frac{1}{4}\left| \varphi ^{M_{d}}_{1}(-{\bf q})\right| ^{2} \nonumber\\
& = & \frac{1}{4}\left| \varphi ^{M_{d}}_{1}(q,\theta )\right| ^{2}
     +\frac{1}{4}\left| \varphi ^{M_{d}}_{0}(q,\theta )\right| ^{2}
     +\frac{1}{4}\left| \varphi ^{M_{d}}_{1}(q,\pi - \theta )\right| ^{2} . \label{label23} 
\end{eqnarray}

\subsection{The Deuteron Eigenvalue Equation in the Helicity Basis}

The deuteron state $|\Psi _{d}^{M_{d}}\rangle$ satisfies the
eigenvalue equation
\begin{equation}
\left( H_{0}-E_{d}+V\right) \left| \Psi _{d}^{M_{d}}\right\rangle =0,
\end{equation}
with $ E_{d} $ being the deuteron binding energy and $V$ the NN potential. 
This eigenvalue equation is projected on 
the basis states $ \left| {\bf q};\hat{q}1\Lambda ;0\right\rangle ^{1a}$ 
introduced in the previous Section. Using
Eq.~(\ref{label2}) for representing $| \Psi _{d}^{M_{d}}\rangle $ one
obtains
\begin{eqnarray}
\lefteqn{\left. ^{^{}}\right. ^{1a}\left\langle {\bf q};\hat{q}1\Lambda ;0\right| \left( H_{0}-E_{d}+V\right) \left| \Psi _{d}^{M_{d}}\right\rangle } \nonumber\\
 &  & \quad \quad = \left. ^{^{}}\right. ^{1a}\left\langle {\bf q};\hat{q}1\Lambda ;0\right| \left( H_{0}-E_{d}\right) \left| \Psi _{d}^{M_{d}}\right\rangle \nonumber \\
 &  & \quad \quad \quad +\left. ^{^{}}\right. ^{1a}\left\langle {\bf q};\hat{q}1\Lambda ;0\right| V\int d{\bf q}'\left\{ \frac{1}{2}\left| {\bf q}';\hat{q}'11;0\right\rangle ^{1a}\varphi ^{M_{d}}_{1}(q',\theta ') + \frac{1}{4}\left| {\bf q}';\hat{q}'10;0\right\rangle ^{1a}\varphi ^{M_{d}}_{0}(q',\theta ')\right\} e^{iM_{d}\phi '}\nonumber \\
 &  & \quad \quad = \left( \frac{q^{2}}{m}-E_{d}\right) \varphi ^{M_{d}}_{\Lambda }(q,\theta )e^{iM_{d}\phi }\nonumber \\
 &  & \quad \quad \quad +\frac{1}{2}\int d{\bf q}'\left. ^{^{}}\right. ^{1a}\left\langle {\bf q};\hat{q}1\Lambda ;0\right| V\left| {\bf q}';\hat{q}'11;0\right\rangle ^{1a}\varphi ^{M_{d}}_{1}(q',\theta ')e^{iM_{d}\phi '}\nonumber \\
 &  & \quad \quad \quad +\frac{1}{4}\int d{\bf q}'\left. ^{^{}}\right. ^{1a}\left\langle {\bf q};\hat{q}1\Lambda ;0\right| V\left| {\bf q}';\hat{q}'10;0\right\rangle ^{1a}\varphi ^{M_{d}}_{0}(q',\theta ')e^{iM_{d}\phi '}\nonumber \\
 &  & \quad \quad = \left( \frac{q^{2}}{m}-E_{d}\right) \varphi ^{M_{d}}_{\Lambda }(q,\theta )e^{iM_{d}\phi }\nonumber \\
 &  & \quad \quad \quad +\int d{\bf q}'\left\{ \frac{1}{2}V_{\Lambda 1}^{110}({\bf q},{\bf q}')\varphi ^{M_{d}}_{1}(q',\theta ') + \frac{1}{4}V_{\Lambda 0}^{110}({\bf q},{\bf q}')\varphi ^{M_{d}}_{0}(q',\theta ')\right\} e^{iM_{d}\phi '} ,
\end{eqnarray}
where $ V_{\Lambda \Lambda '}^{\pi St}({\bf q},{\bf q}')\equiv 
\left. ^{^{}}\right. ^{\pi a}\left\langle {\bf q};\hat{q}S\Lambda ;t\right|
 V\left| {\bf q}';\hat{q}'S\Lambda ';t\right\rangle ^{\pi a} $.
Thus, the eigenvalue equation for deuteron binding energy consists 
in the helicity basis of a set of two coupled integral equations.
\begin{eqnarray}
\lefteqn{\left( \frac{q^{2}}{m}-E_{d}\right) \varphi ^{M_{d}}_{\Lambda }(q,\theta )} \nonumber\\
&  & +\int d{\bf q}'e^{-iM_{d}(\phi -\phi ')}\left\{ \frac{1}{2}V_{\Lambda 1}^{110}({\bf q},{\bf q}')\varphi ^{M_{d}}_{1}(q',\theta ')+\frac{1}{4}V_{\Lambda 0}^{110}({\bf q},{\bf q}')\varphi ^{M_{d}}_{0}(q',\theta ')\right\} = 0 , \label{label1}
\end{eqnarray}
where the index $ \Lambda $ corresponds to 1 or 0. 
Since the wave function components $ \varphi ^{M_{d}}_{\Lambda }(q,\theta ) $ have no
azimuthal dependence, the $ \phi ' $-integral in Eq.~(\ref{label1})
can be carried out independently. Defining
\begin{equation}
v^{\pi St,M_{d}}_{\Lambda \Lambda '}(q,q',\theta,\theta ')\equiv \int ^{2\pi }_{0}d\phi 'e^{-iM_{d}(\phi -\phi ')}V_{\Lambda \Lambda '}^{\pi St}({\bf q},{\bf q}') , \label{label27}
\end{equation}
the coupled integral equations are actually only two-dimensional and
their final form reads
\begin{eqnarray}
\left( \frac{q^{2}}{m}-E_{d}\right) \varphi ^{M_{d}}_{\Lambda }(q,\theta ) & + & \frac{1}{2}\int _{0}^{\infty }dq'\, q'^{2}\int _{-1}^{1}d\cos \theta'\, v^{110,M_{d}}_{\Lambda 1}(q,q',\theta,\theta ')\varphi ^{M_{d}}_{1}(q',\theta ')\nonumber \\
 & + & \frac{1}{4}\int _{0}^{\infty }dq'\, q'^{2}\int _{-1}^{1}d\cos \theta '\,
v^{110,M_{d}}_{\Lambda 0}(q,q',\theta,\theta ')\varphi ^{M_{d}}_{0}(q',\theta ')=0 . \label{label20} 
\end{eqnarray}

The eigenvalue equation, Eq.~(\ref{label20}), is 
consistent with our treatment of the NN continuum of Ref. \cite{1},
where we derived coupled integral 
equations in two variables for the NN scattering equation. 
We would like to add the remark that it is not
necessary to use any specific information about the 
the deuteron, namely spin, isospin and parity.
The set of equations, Eq.~(\ref{label1}), is valid
for arbitrary $ S $, $ \eta _{\pi }$, and $ t $. Any
calculation based on a realistic NN potential will then reveal 
that a solution of the eigenvalue problem
exists only for the well known quantum numbers. 
Therefore, the scheme laid out above 
automatically provides full insight, no a-priori knowledge needs to 
be employed. 
At this level the question of total angular momentum of that bound state is 
undetermined and will be considered in Section III.

\subsection{\label{sec1} Partial Wave Projection of the Deuteron Wave Function}

In this subsection we want to relate $ \left| \Psi
_{d}^{M_{d}}\right\rangle $ as given in Eq.~(\ref{label2}) to the
standard partial wave representation of the deuteron.
The standard representation of the total angular momentum basis
$\left| q(lS)jm;t\right\rangle$ with the normalization
\begin{equation}
\left\langle q'(l'S')j'm'\right. \left| q(lS)jm\right\rangle   =
\frac{\delta (q'-q)}{q'q}\delta _{l'l}\delta _{S'S}\delta _{j'j}\delta
_{m'm} \label{label18}
\end{equation}
is used for projecting the wave function 
$\left| \Psi _{d}^{M_{d}}\right\rangle$.
Again, we use the fixed spin and isospin of the deuteron. The
quantum numbers $ l,\, j,\, m $ remain arbitrary, and we obtain
\begin{eqnarray}
\psi _{l}(q) & \equiv & \left\langle q(l1)jm;0\right| \left. \Psi _{d}^{M_{d}}\right\rangle \nonumber \\
 & = & \frac{1}{2}\int d{\bf q}'\left\langle q(l1)jm;0\right| \left. {\bf q}';\hat{q}'11;0\right\rangle ^{1a}\varphi ^{M_{d}}_{1}(q',\theta')e^{iM_{d}\phi '}\nonumber \\
 &  & +\frac{1}{4}\int d{\bf q}'\left\langle q(l1)jm;0\right| \left. {\bf
q}';\hat{q}'10;0\right\rangle ^{1a}\varphi ^{M_{d}}_{0}(q',\theta')e^{iM_{d}\phi '} . \label{label4} 
\end{eqnarray}
Recalling the explicit representation of the helicity state \cite{1},
\begin{eqnarray}
\left| {\bf q};\hat{q}S\Lambda \right\rangle  & = & \left| {\bf q}\right\rangle \left| \hat{q}S\Lambda \right\rangle \nonumber \\
 & = & \sum _{ljm}\left| q(lS)jm\right\rangle \sum _{\mu }C(lSj;\mu ,m-\mu ,m)
Y_{l\mu }^{\ast }(\hat{q})e^{-i(m-\mu )\phi }d^{S}_{m-\mu ,\Lambda
}(\theta),
\end{eqnarray}
and using Eq.~(\ref{label3}), the scalar product of partial wave and helicity
basis can be worked out as
\begin{eqnarray}
\left\langle q'(lS')jm;t'\right| \left. {\bf q};\hat{q}S\Lambda ;t\right\rangle ^{\pi a} & = & \left( \left\langle q'(lS')jm\right| \left. {\bf q}\right\rangle \left| \hat{q}S\Lambda \right\rangle +\eta _{\pi }\left\langle q'(lS')jm\right| \left. -{\bf q}\right\rangle \left| \hat{q}S\Lambda \right\rangle \right) \left\langle t'\right| \left. t\right\rangle \nonumber \\
 & = & \frac{\delta (q'-q)}{q'q}\delta _{S'S}\delta _{t't}\nonumber \\
 &  & \sum _{\mu }C(lSj;\mu ,m-\mu ,m)\left( Y_{l\mu }^{\ast }(\hat{q})+\eta _{\pi }Y_{l\mu }^{\ast }(-\hat{q})\right) e^{-i(m-\mu )\phi }d^{S}_{m-\mu ,\Lambda }(\theta )\nonumber \\
 & = & \left( 1+\eta _{\pi }(-)^{l}\right) \frac{\delta (q'-q)}{q'q}\delta _{S'S}\delta _{t't}\nonumber \\
 &  & \sum _{\mu }C(lSj;\mu ,m-\mu ,m)Y_{l\mu }^{\ast }(\hat{q})e^{-i(m-\mu )\phi }d^{S}_{m-\mu ,\Lambda }(\theta )\nonumber \\
 & = & \left( 1+\eta _{\pi }(-)^{l}\right) \sqrt{\frac{2l+1}{4\pi }}\frac{\delta (q'-q)}{q'q}\delta _{S'S}\delta _{t't}e^{-im\phi }\nonumber \\
 &  & \sum _{\mu }C(lSj;\mu ,m-\mu ,m)d^{l}_{\mu 0}(\theta )d^{S}_{m-\mu ,\Lambda }(\theta )\nonumber \\
 & = & \left( 1+\eta _{\pi }(-)^{l}\right) \frac{\delta (q'-q)}{q'q}\delta _{S'S}\delta _{t't}\nonumber\\
&  & \sqrt{\frac{2l+1}{4\pi }} e^{-im\phi } C(lSj;0\Lambda \Lambda
)d^{j}_{m\Lambda }(\theta ).
\end{eqnarray}
Here we used the relation
\begin{equation}
Y^{\ast }_{l\mu }(\hat{q})e^{i\mu \phi }=\sqrt{\frac{2l+1}{4\pi }}d^{l}_{\mu 0}(\theta )
\end{equation}
 together with an addition theorem for Wigner's D-functions
\begin{equation}
\sum _{\mu }C(lSj;\mu ,m-\mu ,m)d^{l}_{\mu 0}(\theta )d^{S}_{m-\mu
,\Lambda }(\theta )=C(lSj;0\Lambda \Lambda )d^{j}_{m\Lambda }(\theta ).
\end{equation}
Hence, the projection of the deuteron state on the partial wave basis as defined
in Eq.~(\ref{label4}) is given by
\begin{eqnarray}
\psi _{l}(q) & = & \left( 1+(-)^{l}\right) \sqrt{\frac{2l+1}{4\pi }}\int _{0}^{2\pi }d\phi 'e^{-i(m-M_{d})\phi '}\nonumber \\
 &  & \int _{-1}^{1}d\cos \theta '\left\{ \frac{1}{2}C(l1j;011)d^{j}_{m1}(\theta ')\varphi ^{M_{d}}_{1}(q,\theta ') 
 + \frac{1}{4}C(l1j;000)d^{j}_{m0}(\theta ')\varphi ^{M_{d}}_{0}(q,\theta ')\right\} \nonumber \\
 & = & \left( 1+(-)^{l}\right) \sqrt{\pi (2l+1)}\delta _{mM_{d}}\nonumber \\
 &  & \int _{-1}^{1}d\cos \theta '\left\{ \frac{1}{2}C(l1j;011)d^{j}_{m1}(\theta ')\varphi ^{M_{d}}_{1}(q,\theta ') 
 + \frac{1}{4}C(l1j;000)d^{j}_{m0}(\theta ')\varphi ^{M_{d}}_{0}(q,\theta ')\right\} 
\end{eqnarray}
This projection exists only for $ m=M_{d} $ and even $l$, 
as enforced by the even deuteron parity, and one obtains
\begin{eqnarray}
\psi _{l}(q) & = & 2\sqrt{\pi (2l+1)}\nonumber \\
 &  & \int _{-1}^{1}d\cos \theta '\left\{ \frac{1}{2}C(l1j;011)d^{j}_{M_{d}1}(\theta ')\varphi ^{M_{d}}_{1}(q,\theta ')
 + \frac{1}{4}C(l1j;000)d^{j}_{M_{d}0}(\theta ')\varphi ^{M_{d}}_{0}(q,\theta ')\right\} \label{label15} 
\end{eqnarray}
At this point the remaining properties for the projections $ \psi _{l}(q) $
can be found by explicit calculation. In other words 
the fact that $l$ has to be 0 and 2, and consequently $j$=1, 
has to be inferred numerically from the solution of Eq.~(\ref{label15}). 
Of course an analytical
investigation can be added once we adopt the deuteron wave function with the
well established analytical angular behavior (see Section~\ref{sec5}).

\subsection{\label{sec4}Explicit Solution of the Deuteron Eigenvalue
Equation}

The numerical solution of the set of two coupled eigenvalue equations Eq.~(\ref{label20}) poses 
no specific difficulty. We employ an iterative, Lanczo's type technique \cite{5,6}, which 
provides both, eigenvalue and eigenvector. Following Ref. \cite{7} 
the method is modified  to avoid unphysical solutions  corresponding to 
bound states in the repulsive core region. 
All numerical calculations are carried out with 
the Bonn-B potential \cite{bonnb}.

For the discretization of Eq.~(\ref{label20}) we employ Gaussian grids.
The $\phi '$ integration over the potential,  Eq.~(\ref{label27}), 
needs only 10 quadrature points, whereas 
the $\cos \theta '$ integration requires at least 32 grid points and the $q'$ 
integration 
72 grid points, depending on the desired accuracy. The $q'$ integration  interval can 
safely be cut off at 30 fm$^{-1}$. Using these numerical parameters,  we obtain the deuteron 
binding energy 2.224 MeV.

In Fig.~\ref{fig1} the two deuteron wave function components 
$ \varphi ^{M_{d}}_{\Lambda }(q,\theta) $, $\Lambda =0,1$, are shown 
for $ M_{d}=0 $ as functions of $ q $ and $ \cos \theta  $. Both drop
quickly with increasing  relative momentum between the two nucleons 
inside the deuteron.
The wave function $ \varphi ^{0}_{0}(q,\theta) $ shows a cosine-like behavior 
indicated by the straight line at $q=0$ connecting the highest
point at $ \theta =0 $ with the lowest point at $ \theta =180^{o} $ through
zero at $ \theta =90^{o} $. This cosine-like behavior is confirmed to be
true when considering the analytical angular behavior of the wave functions
in Section \ref{sec3}. The function $ \varphi ^{0}_{1}(q,\theta)$, in
contrast, displays a sine-like behavior, which also will prove to be
its correct analytical form.
It peaks at $ \theta =90^{o} $ and vanishes at $ \theta =0 $
and $ 180^{o} $. The maximum of $ \varphi ^{0}_{0}(q,\theta) $ is larger than
that of $ \varphi ^{0}_{1}(q,\theta) $.

For $ M_{d}=1 $ and $ -1 $ the wave function components $ \varphi ^{M_{d}}_{\Lambda }(q,\theta) $
are shown as functions of $ q $ and $ \cos \theta  $ in Fig.~\ref{fig2}.
They also drop quickly as the relative momentum between the two nucleons inside the
deuteron increases. In the upper part of the figure we see that both $ \varphi ^{1}_{0}(q,\theta) $
and $ \varphi ^{-1}_{0}(q,\theta) $ vanish at $ \theta =0 $ and $ 180^{o} $
but have a  different sign for the other $\theta$-values. At $ \theta =90^{o} $
$ \varphi ^{1}_{0}(q,\theta) $ reaches its minimum whereas $ \varphi ^{-1}_{0}(q,\theta) $
reaches its maximum. 
In the lower part of the figure we see that $ \varphi ^{1}_{1}(q,\theta) $
peaks at $ \theta =0 $ and vanishes at $ \theta =180^{o} $. On the contrary
$ \varphi ^{-1}_{1}(q,\theta) $ peaks at $ \theta =180^{o} $ and vanishes at
$ \theta =0 $. This angular behavior of $ \varphi ^{1}_{\Lambda }(q,\theta) $
and $ \varphi ^{-1}_{\Lambda }(q,\theta) $ suggests a relation between
the two functions. This relation is explicitly given 
in Section \ref{sec3}. 
For $ M_{d}=1 $ and $ -1 $ the maximum of 
$\varphi ^{M_{d}}_{1}(q,\theta)$ is larger than that of $\varphi^{M_{d}}_{0}(q,\theta)$.

In Fig.~\ref{fig3} the deuteron densities $\rho ^{M_{d}}({\bf q})$ as given in 
Eq.~(\ref{label23}) for $M_{d}=0$ (top row)  and
$M_{d}=1$ (bottom row)  are shown. On the left side the two densities
are displayed as functions
of $ q $ and $ \cos \theta$,  and on the right side 
as  functions of the Cartesian projections of $\bf q$,  $q_x$ and $q_z$. 
Since the wave functions are invariant under rotations around the
z-axis, we  show a vertical cut through the x-z-plane.
The densities $ \rho ^{0}({\bf q}) $ and $ \rho ^{1}({\bf q}) $ have 
uniform angular distributions
with respect to the azimuthal angle $\theta$ for the momentum range shown, and thus
the equidensity surfaces as function of the momentum between the two
nucleons have a spherical shape. The densities are largest at small relative momentum ${\bf
q}$. Though not shown here the deuteron density for $ M_{d}=-1 $
also has a similar shape. In all Figs. 1-3 the densities have their maximum at $q=0$, indicating that 
the two nucleons being at rest with respect to each other is the most probable configuration for the deuteron.
 
Using Eq.~(\ref{label15}) we extract the usual $S$ and $D$ wave components. 
They agree very well with the ones obtained from standard partial wave calculations.


\section{Formulation II}

\subsection{Deuteron Wave Function in Operator Form}

In order to study the different spin orientations of the two nucleons in the
deuteron in relation to the vector of relative momentum a representation 
of the deuteron wave function in operator form is ideal. It is also desirable to derive another 
set of two coupled, one-dimensional equations, in the basis of total helicity. 

In terms of the partial wave basis states given in Eq.~(\ref{label18}) the
deuteron state has the well known form
\begin{equation}
\left| \Psi _{d}^{M_{d}}\right\rangle =\left| t\right\rangle \sum _{l=0,2}\int
_{0}^{\infty }dq\, q^{2}\left| q(l1)1M_{d}\right\rangle \psi _{l}(q).
\end{equation}
Here $ \left| t\right\rangle $ indicates the isospin, which is 0 for the deuteron. 
The explicit reference to t will be omitted in the following considerations. 
Again, we would like to point out that we work in the basis of total helicity,
and thus our final expressions will differ from the ones given in Ref. \cite{alzetta}. 
Carrying out
 the angular momentum expansion explicitly and using 
$\left\langle \hat{q}\right| \left. l\mu \right\rangle =Y_{l\mu }(\hat{q}) $ one
obtains
\begin{eqnarray}
\Psi _{d}^{M_{d}}({\bf q}) & = & \left| 1M_{d}\right\rangle \frac{1}{\sqrt{4\pi }}\psi _{0}(q)\nonumber \\
 &  & + \left\{ \left| 11\right\rangle C(211;M_{d}-1,1M_{d})Y_{2,M_{d}-1}(\hat{q})\right. \nonumber \\
 &  & +\left| 10\right\rangle C(211;M_{d}0M_{d})Y_{2M_{d}}(\hat{q})\nonumber \\
 &  & +\left. \left| 1-1\right\rangle
C(211;M_{d}+1,-1,M_{d})Y_{2,M_{d}+1}(\hat{q})\right\} \psi _{2}(q).
\end{eqnarray}
Inserting the explicit expressions for the Clebsch-Gordon coefficients \cite{9}
leads to
\begin{eqnarray}
\Psi _{d}^{M_{d}}({\bf q}) & = & \left| 1M_{d}\right\rangle \frac{1}{\sqrt{4\pi }}\psi _{0}(q)\nonumber \\
 &  & + \left\{ \left| 11\right\rangle \sqrt{\frac{(2-M_{d})(3-M_{d})}{20}}Y_{2,M_{d}-1}(\hat{q})\right. \nonumber \\
 &  & -\left| 10\right\rangle \sqrt{\frac{(2-M_{d})(2+M_{d})}{10}}Y_{2M_{d}}(\hat{q})\nonumber \\
 &  & +\left. \left| 1-1\right\rangle
\sqrt{\frac{(2+M_{d})(3+M_{d})}{20}}Y_{2,M_{d}+1}(\hat{q})\right\} \psi _{2}(q).
\end{eqnarray}

\noindent
Now we would like to express the wave function in a simple way such that
\begin{equation}
\label{label8}
\Psi _{d}^{M_{d}}({\bf q})=\left\{ c_{0}\psi _{0}(q)+c_{2}\psi _{2}(q)\right\} \left|
1M_{d}\right\rangle,
\end{equation}
and where $ c_{0} $ and $ c_{2} $ are operators acting on the deuteron spin
state $ \left| 1M_{d}\right\rangle  $. For this purpose we choose as  example
$ M_{d}=1 $ which yields
\begin{eqnarray}
\Psi _{d}^{1}({\bf q}) & = & \left| 11\right\rangle \frac{1}{\sqrt{4\pi }}\psi _{0}(q)
+\left\{ \left| 11\right\rangle \sqrt{\frac{1}{10}}Y_{20}(\hat{q})
-\left| 10\right\rangle \sqrt{\frac{3}{10}}Y_{21}(\hat{q})
+\left| 1-1\right\rangle \sqrt{\frac{3}{5}}Y_{22}(\hat{q})\right\} \psi _{2}(q)\nonumber \\
& = & \left| 11\right\rangle \frac{1}{\sqrt{4\pi }}\psi _{0}(q)
+\left\{ \left| 11\right\rangle (q_{0}^{2}+q_{1}q_{-1})-\left| 10\right\rangle
3q_{0}q_{1}+\left| 1-1\right\rangle 3q_{1}^{2}\right\}
\frac{1}{2q^{2}}\sqrt{\frac{1}{2\pi }} \psi _{2}(q) . \label{label7} 
\end{eqnarray}
In the last step we expressed the spherical harmonic functions in terms of the spherical
components of the momentum, $ q_{1},\, q_{0} $ and $ q_{-1} $ \cite{9}.
Since the state $ \left| 1M_{d}\right\rangle  $ in Eq.~(\ref{label8}) has
already the correct transformation property under rotation of the deuteron state,
the operators $ c_{0} $ and $ c_{2} $ must be scalars under rotation.
Those scalars have to be formed out of the spherical components of
${\ffat \sigma }(1)$ and ${\ffat \sigma }(2) $ which at the same
time will connect the given states $ \left| 1-1\right\rangle ,\, \left|10\right\rangle$
and $\left| 11\right\rangle$ to $\left| 1M_{d}\right\rangle$. Therefore
we consider
\begin{eqnarray}
{\ffat \sigma }(1)\cdot {\bf q}\, {\ffat \sigma }(2)\cdot {\bf q}\left| 11\right\rangle  & = & \left( \sigma _{0}(1)q_{0}-\sigma _{1}(1)q_{-1}-\sigma _{-1}(1)q_{1}\right) \left| \frac{1}{2}\frac{1}{2}\right\rangle \nonumber \\
 &  & \left( \sigma _{0}(2)q_{0}-\sigma _{1}(2)q_{-1}-\sigma _{-1}(2)q_{1}\right) \left| \frac{1}{2}\frac{1}{2}\right\rangle \nonumber \\
 & = & q_{0}^{2}\left| 11\right\rangle -2q_{0}q_{1}\left| 10\right\rangle
+2q_{1}^{2}\left| 1-1\right\rangle . 
\end{eqnarray}
The $ l=0 $ admixture can be projected out by subtracting $\frac{1}{3}q^{2}$,
which leads to
\begin{equation}
\left( {\ffat \sigma }(1)\cdot {\bf q}\, {\ffat \sigma }(2)\cdot {\bf
q}-\frac{1}{3}q^{2}\right) \left| 11\right\rangle = \frac{2}{3}\left(
q_{0}^{2}+q_{1}q_{-1}\right) \left| 11\right\rangle -2q_{0}q_{1}\left|
10\right\rangle +2q_{1}^{2}\left| 1-1\right\rangle. \label{label6} 
\end{equation}
A comparison to the terms in Eq.~(\ref{label7}) reveals that $ \Psi _{d}^{1}({\bf q}) $
can be written as
\begin{eqnarray}
\Psi _{d}^{1}({\bf q}) & = & \left\{ \frac{1}{\sqrt{4\pi }}\psi _{0}(q)+\left[ {\ffat \sigma }(1)\cdot {\bf q}\, {\ffat \sigma }(2)\cdot {\bf q}-\frac{1}{3}q^{2}\right] \frac{3}{4q^{2}}\sqrt{\frac{1}{2\pi }}\psi _{2}(q)\right\} \left| 11\right\rangle \nonumber \\
 & = & \left\{ \bar{\psi }_{0}(q)+\left[ {\ffat \sigma }(1)\cdot {\bf q}\, {\ffat
\sigma }(2)\cdot {\bf q}-\frac{1}{3}q^{2}\right] \bar{\psi }_{2}(q)\right\}
\left| 11\right\rangle , \label{label5} 
\end{eqnarray}
where 
\begin{eqnarray}
\bar{\psi }_{0}(q) & \equiv  & \frac{1}{\sqrt{4\pi }}\psi _{0}(q)\\
\bar{\psi }_{2}(q) & \equiv  & \frac{3}{4q^{2}}\frac{1}{\sqrt{2\pi }}\psi _{2}(q)
\end{eqnarray}

\noindent
We denote the expression in Eq.~(\ref{label5}) as `operator form' of the deuteron
wave function in momentum space. A corresponding expression in coordinate
space can be found in Ref.~\cite{ericson}. In a fashion similar to the above
derivation,
 one can show that the form given in Eq.~(\ref{label5}) is also valid
for $ M_{d}=0 $ and $ -1 $. Hence, the deuteron wave function in operator
form is given in momentum space  as 
\begin{equation}
\label{label10}
\Psi _{d}^{M_{d}}({\bf q})=\left\{ \bar{\psi }_{0}(q)+\left[ {\ffat \sigma
}(1)\cdot {\bf q}\, {\ffat \sigma }(2)\cdot {\bf q}-\frac{1}{3}q^{2}\right]
\bar{\psi }_{2}(q)\right\} \left| 1M_{d}\right\rangle .
\end{equation}
Here the positive parity is manifest, 
since $ \Psi _{d}^{M_{d}}({\bf q})=\Psi _{d}^{M_{d}}(-{\bf q}) $. 
It is a straightforward algebra to work out the normalization of 
$\left| \Psi _{d}^{M_{d}}\right\rangle $ as given in Eq.~(\ref{label10}) and one
obtains
\begin{eqnarray}
\left\langle \Psi _{d}^{M_{d}}\right| \left. \Psi _{d}^{M_{d}}\right\rangle & = & \int d{\bf q} \left\langle 1M_{d}\right| \left\{ \bar{\psi }_{0}(q)+\left[ {\ffat \sigma }(1)\cdot {\bf q}\, {\ffat \sigma }(2)\cdot {\bf q}-\frac{1}{3}q^{2}\right] \bar{\psi }_{2}(q)\right\} \nonumber \\
 &  & \left\{ \bar{\psi }_{0}(q)+\left[ {\ffat \sigma }(1)\cdot {\bf q}\, {\ffat \sigma }(2)\cdot {\bf q}-\frac{1}{3}q^{2}\right] \bar{\psi }_{2}(q)\right\} \left| 1M_{d}\right\rangle \nonumber \\
 & = & 4\pi \int _{0}^{\infty }dq\, q^{2}\left\{ \bar{\psi }_{0}^{2}(q)+\frac{8}{9}q^{4}\bar{\psi }_{2}^{2}(q)\right\} \nonumber \\
 & = & \int _{0}^{\infty }dq\, q^{2}\left\{ \psi _{0}^{2}(q)+\psi
_{2}^{2}(q)\right\}. \label{label9} 
\end{eqnarray}
The last form is the standard normalization of the deuteron wave function in
terms of partial wave components. In arriving at this result we used that
\begin{eqnarray}
\left\langle 1M_{d}\right| {\ffat \sigma }(1)\cdot {\bf q}\, {\ffat \sigma }(2)\cdot {\bf q}\left| 1M_{d}\right\rangle  & = & \left\{ \begin{array}{ll}
q_{0}^{2} & ,\, M_{d}=\pm 1\\
q^{2}-2q_{0}^{2} & ,\, M_{d}=0
\end{array}\right. 
\end{eqnarray}
and
\begin{eqnarray}
\int d \hat{q}\, \left\langle 1M_{d}\right| {\ffat \sigma }(1)\cdot {\bf q}\,
{\ffat \sigma }(2)\cdot {\bf q}\left| 1M_{d}\right\rangle  & = & \frac{4\pi
}{3}q^{2} .
\end{eqnarray}

\noindent
As we shall show in Section \ref{SEC2} based on the form given in
 Eq.~(\ref{label10}) one can express the
angular dependencies of all possible spin orientations in the deuteron analytically.

\subsection{\label{sec3} Analytical Angular Behavior of the Deuteron Wave
Function and the Deuteron Eigenvalue Equation}

With the operator form Eq.~(\ref{label10}) at hand we revisit the deuteron
wave function component in the helicity basis as given in Eq.~(\ref{label11}):
\begin{eqnarray}
\varphi ^{M_{d}}_{\Lambda }({\bf q}) & \equiv  & \left. ^{^{}}\right. ^{1a}\left\langle {\bf q};\hat{q}1\Lambda ;0\right| \left. \Psi _{d}^{M_{d}}\right\rangle \nonumber \\
 & = & \left\langle \hat{q}1\Lambda \right| \left( \left\langle {\bf q}\right| +\left\langle -{\bf q}\right| \right) \left. \Psi _{d}^{M_{d}}\right\rangle \nonumber \\
 & = & 2 \left\langle \hat{q}1\Lambda \right| \Psi _{d}^{M_{d}}({\bf
q}) \nonumber \\
 & = & 2\left\langle \hat{q}1\Lambda \right| \left\{ \bar{\psi }_{0}(q)+\left[ {\ffat
\sigma }(1)\cdot {\bf q}\, {\ffat \sigma }(2)\cdot {\bf q}-\frac{1}{3}q^{2}\right]
\bar{\psi }_{2}(q)\right\} \left| 1M_{d}\right\rangle .
\end{eqnarray}
The operator $ {\ffat \sigma }(1)\cdot {\bf q}\, {\ffat \sigma }(2)\cdot {\bf q} $
can be expressed in terms of the total helicity $ {\bf S}\cdot {\bf q} $ as
\begin{equation}
{\ffat \sigma }(1)\cdot {\bf q}\, {\ffat \sigma }(2)\cdot {\bf q}=2({\bf S}\cdot {\bf
q})^{2}-q^{2},
\end{equation}
where  ${\bf S}=\frac{1}{2}\left( {\ffat \sigma }(1)+{\ffat \sigma }(2)\right)$.
Therefore, the helicity wave function component is given by
\begin{eqnarray}
\varphi ^{M_{d}}_{\Lambda }({\bf q}) & = & 2\left\langle \hat{q}1\Lambda \right| \left\{ \bar{\psi }_{0}(q)+\left[ 2({\bf S}\cdot {\bf q})^{2}-\frac{4}{3}q^{2}\right] \bar{\psi }_{2}(q)\right\} \left| 1M_{d}\right\rangle \nonumber \\
 & = & 2\left\{ \bar{\psi }_{0}(q)+\left[ 2\Lambda ^{2}-\frac{4}{3}\right] q^{2}\bar{\psi }_{2}(q)\right\} \left\langle \hat{q}1\Lambda \right| \left. 1M_{d}\right\rangle \nonumber \\
 & = & 2\left\{ \bar{\psi }_{0}(q)+\left[ 2\Lambda ^{2}-\frac{4}{3}\right] q^{2}\bar{\psi }_{2}(q)\right\} D^{1\ast }_{M_{d}\Lambda }(\phi \theta 0)\nonumber \\
 & = & 2\left\{ \bar{\psi }_{0}(q)+\left[ 2\Lambda ^{2}-\frac{4}{3}\right]
q^{2}\bar{\psi }_{2}(q)\right\} e^{iM_{d}\phi }d^{1}_{M_{d}\Lambda }(\theta ).
\end{eqnarray}
This shows that the angular behavior of the wave function component is given by  
$e^{iM_{d}\phi }d^{1}_{M_{d}\Lambda }(\theta )$, where the 
d-matrix is explicitly given as \cite{9}
\begin{eqnarray}
\label{label30}
d_{M_{d}\Lambda }^{1}(\theta ) & = & \left( \begin{array}{ccc}
\frac{1+\cos \theta }{2} & -\frac{\sin \theta }{\sqrt{2}} & \frac{1-\cos \theta }{2}\\
\frac{\sin \theta }{\sqrt{2}} & \cos \theta  & -\frac{\sin \theta }{\sqrt{2}}\\
\frac{1-\cos \theta }{2} & \frac{\sin \theta }{\sqrt{2}} & \frac{1+\cos \theta }{2}
\end{array}\right) .
\end{eqnarray}

\noindent
Finally, we define an angle independent function $ \Phi _{\Lambda }(q)$ 
via
\begin{eqnarray}
\Phi _{\Lambda }(q) & \equiv  & \bar{\psi }_{0}(q)+\left[ 2\Lambda ^{2}-\frac{4}{3}\right] q^{2}\bar{\psi }_{2}(q)\nonumber \\
 & = & \frac{1}{\sqrt{4\pi }}\psi _{0}(q)+\left[ 3\Lambda ^{2}-2\right]
\frac{1}{\sqrt{8\pi }}\psi _{2}(q), \label{label16} 
\end{eqnarray}
so that the deuteron wave function component can be expressed as
\begin{eqnarray}
\varphi ^{M_{d}}_{\Lambda }({\bf q}) & = & 2\Phi _{\Lambda }(q)e^{iM_{d}\phi }d^{1}_{M_{d}\Lambda }(\theta )\nonumber\\
& \equiv & \varphi ^{M_{d}}_{\Lambda }(q,\theta)e^{iM_{d}\phi } . \label{label12}
\end{eqnarray}
Here $\varphi ^{M_{d}}_{\Lambda }(q,\theta)$ are the wave function components which we
determined previously numerically and which are displayed in
Figs.~\ref{fig1}-\ref{fig2}.


Employing the above given form of the deuteron wave function,
we can derive a one-dimensional eigenvalue equation, starting from 
the eigenvalue equation for 
$\varphi ^{M_{d}}_{\Lambda }(q,\theta )=\varphi ^{M_{d}}_{\Lambda }({\bf q})
\exp (-iM_d\phi)$.
Inserting Eq.~(\ref{label12}) into Eq.~(\ref{label1}) gives
\begin{eqnarray}
\lefteqn {\left( \frac{q^{2}}{m}-E_{d}\right) \Phi _{\Lambda }(q)d^{1}_{M_{d}\Lambda }(\theta )} &  & \nonumber \\
 &  & +\int d{\bf q}'e^{-iM_{d}(\phi -\phi ')}\left\{ \frac{1}{2}V_{\Lambda
1}^{110}({\bf q},{\bf q}')\Phi _{1}(q')d^{1}_{M_{d}1}(\theta ') +
\frac{1}{4}V_{\Lambda 0}^{110}({\bf q},{\bf q}')\Phi
_{0}(q')d^{1}_{M_{d}0}(\theta ')\right\} =0, \label{label14} 
\end{eqnarray}
an equation, which is valid for any direction $ \theta  $.
Choosing $ \hat{q}=\hat{z} $ simplifies the equation, since the azimuthal
dependencies of the potential can be factored out as
\begin{equation}
V_{\Lambda \Lambda '}^{\pi St}(q\hat{z},{\bf q}')\equiv e^{i\Lambda (\phi
-\phi ')}V_{\Lambda \Lambda '}^{\pi St}(q,q',\theta ').
\end{equation}
The d-matrix in the first term gives  $ \delta _{M_{d}\Lambda } $,
and the $\phi '$ integration requires  $\Lambda$ to be equal $M_d$, leading 
to 
\begin{eqnarray}
\lefteqn {\left( \frac{q^{2}}{m}-E_{d}\right) \Phi _{M_{d}}(q)} \nonumber\\
 &  & +\pi \int _{0}^{\infty }dq'\, q'^{2}\int _{-1}^{1}d \cos \theta' 
\left\{V_{M_{d}1}^{110}(q,q',\theta ')\Phi _{1}(q')d^{1}_{M_{d}1}(\theta ') +
\frac{1}{2}V_{M_{d}0}^{110}(q,q',\theta ')\Phi _{0}(q')d^{1}_{M_{d}0}(\theta
')\right\} =0 .  \nonumber\\ \label{label19} 
\end{eqnarray}
Choosing $M_d= 1$ and $0$ leads to a closed system of two coupled equations for 
the amplitudes $\Phi_1$ and $\Phi_0$.
The $\cos \theta ' $ integration  can be worked out independent of the amplitudes 
$\Phi_\Lambda (q')$, so that   Eq.~(\ref{label19}) is in fact 
a set of two coupled equations in one variable, namely $q$:
\begin{equation}
\left( \frac{q^{2}}{m}-E_{d}\right) \Phi _{M_{d}}(q)+\pi \int _{0}^{\infty
}dq'\, q'^{2}\left\{ V_{M_{d}1}^{110}(q,q')\Phi
_{1}(q')+\frac{1}{2}V_{M_{d}0}^{110}(q,q')\Phi _{0}(q')\right\} =0 , \label{label21} 
\end{equation}
with
\begin{equation}
V_{M_{d}\Lambda '}^{110}(q,q')\equiv \int _{-1}^{1}d\cos \theta '\, V_{M_{d}\Lambda
'}^{110}(q,q',\theta ')d^{1}_{M_{d}\Lambda '}(\theta ') . \label{label28}
\end{equation}

The set of two coupled eigenvalue equations (\ref{label21}) can be easily
solved using the same method as described 
in Section \ref{sec4}. The Gaussian grids for 
the $q'$ integration and the $\cos \theta '$ integration 
in Eq.~(\ref{label28}) are taken to be the same, and we obtain the same value for
the deuteron binding energy, $E_d$=~2.224 MeV. 
The solutions $ \Phi _{0}(q) $ and $ \Phi _{1}(q) $ are displayed in Fig.~\ref{fig4}. 
This figure shows that both functions are of the same magnitude for $q=0$, and
both drop by about one order of magnitude within  
$q$ of $\approx 200$~MeV/c. $\Phi _{1}(q)$ has its first node already for
$q\approx 300$~MeV/c, while the first node of $\Phi _{0}(q)$ occurs for 
$q\approx 800$~MeV/c. In general, the magnitude of $\Phi _{0}(q)$ falls off
slightly slower than the one for $\Phi _{1}(q)$ as function of $q$. 

In Figs.~\ref{fig1} and \ref{fig2} the wave function components 
$\varphi^{M_d}_\Lambda (q,\theta)$  are obtained 
from numerically solving Eq.~(\ref{label20}).  With the help of Eqs.~(\ref{label12}) 
and (\ref{label30}) we can express their angular behavior as
\begin{eqnarray}
M_{d}=0 & : & \varphi ^{0}_{0}(q,\theta )=2\Phi _{0}(q)\cos \theta \label{eq3.26} \\
 &  & \varphi ^{0}_{1}(q,\theta )=\sqrt{2}\Phi _{1}(q)\sin \theta \\
M_{d}=1 & : & \varphi ^{1}_{0}(q,\theta )=-\sqrt{2}\Phi _{0}(q)\sin \theta \\
 &  & \varphi ^{1}_{1}(q,\theta )=\Phi _{1}(q)(1+\cos \theta ) \label{eq3.29} .
\end{eqnarray}
Obviously, the angular behavior extracted numerically
agrees with the analytical one.

We mentioned in Section \ref{sec4} that the maximum of $ \varphi ^{M_d}_{\Lambda}(q,\theta ) $ with
$ \Lambda = M_d $ is larger than that with $ \Lambda \neq M_d $. Eqs.~(\ref{eq3.26})-(\ref{eq3.29}) show 
that the ratio $| \varphi ^{M_d}_{M_d}(q,\theta ) _{max} / \varphi ^{M_d}_{\Lambda \neq M_d}(q,\theta ) _{max}
|$ is exactly $ \sqrt{2} $. This can be understood as follow. According to Eq.~(\ref{label16}) the component 
$ \varphi ^{M_d}_{\Lambda}(q,\theta ) $ is determined for small $q$ dominantly by $ \psi _{0}(q) $, 
i.e. the $S$ wave.

The analytical angular behavior of the deuteron densities 
given in Eq.~(\ref{label23}) can now  easily be derived. 
For $ M_{d}=0,1$ 
we find
\begin{eqnarray}
\rho ^{0}({\bf q}) & = & \Phi ^{2}_{1}(q)\sin ^{2}\theta +\Phi ^{2}_{0}(q)\cos ^{2}\theta \label{eq3.30} \\
\rho ^{1}({\bf q}) & = & \frac{1}{2}\Phi ^{2}_{1}(q)(1+\cos ^{2}\theta )
                        +\frac{1}{2}\Phi ^{2}_{0}(q)\sin ^{2}\theta . \label{eq3.31}
\end{eqnarray}
From these expressions we can deduce that $ \rho ^{0}({\bf q}) $ and $ \rho ^{1}({\bf q}) $ are only
perfect spheres for small $q$, where $\Phi _{0}(q)$ and $\Phi _{1}(q)$ are
almost identical. For larger momenta the spheres are deformed according the
ratio  $ \left| \Phi _{0}(q)/\Phi _{1}(q)\right| $.

\subsection{\label{sec5}Relation to the Conventional  Partial Wave Representation}

Before completing the considerations about the analytic behavior of the
angular behavior of the deuteron wave function in the helicity basis, we want
to make contact with the standard representation of the deuteron wave function. 
In Section \ref{sec1} we derived the projection of the deuteron state on
the partial wave basis. We ended up with Eq.~(\ref{label15}) and left
the remaining quantum numbers $j$ and $l$ to be determined numerically. 
The wave function components
$\varphi_\Lambda (q,\theta)$ together with their analytical angle behavior allows
to calculate the projection and to determine
the remaining conditions for their existence. Inserting
Eq.~(\ref{label12})
into Eq.~(\ref{label15}) yields
\begin{eqnarray}
\psi _{l}(q) & = & 2\sqrt{\pi (2l+1)}\nonumber \\
 &  & \int _{-1}^{1}d\cos \theta '\left\{ C(l1j;011)d^{j}_{M_{d}1}(\theta ')\Phi
_{1}(q)d^{1}_{M_{d}1}(\theta ')
+ \frac{1}{2}C(l1j;000)d^{j}_{M_{d}0}(\theta ')\Phi
_{0}(q)d^{1}_{M_{d}0}(\theta ')\right\} \nonumber \\
 & = & \frac{4}{3}\delta _{j1}\sqrt{\pi (2l+1)}\left\{ C(l1j;011)\Phi
_{1}(q)+\frac{1}{2}C(l1j;000)\Phi _{0}(q)\right\}.
\end{eqnarray}
Here we use the orthogonality property of the d-matrix
\begin{equation}
\int _{-1}^{1}d\cos \theta \, d^{j_{1}}_{\mu _{1}m_{1}}(\theta )d^{j_{2}}_{\mu
_{2}m_{2}}(\theta )=\frac{2}{2j_{1}+1}\delta _{j_{1}j_{2}}\delta _{\mu _{1}\mu
_{2}}\delta _{m_{1}m_{2}}.
\end{equation}
The projection exists only for  a total angular momentum $ j=1 $. Furthermore,
the Clebsch-Gordon coefficients allow only $ l=0 $ and $ l=2 $ and we obtain
explicitly for the $S$ and $D$ wave
\begin{eqnarray}
\psi _{0}(q) & = & \frac{2}{3}\sqrt{\pi }\left\{ 2\Phi _{1}(q)+\Phi
_{0}(q)\right\}\label{label25} \\
\psi _{2}(q) & = & \frac{2}{3}\sqrt{2\pi }\left\{ \Phi _{1}(q)-\Phi
_{0}(q)\right\} , \label{label26}
\end{eqnarray}
which is consistent with Eq.~(\ref{label16}). We extracted the $S$ and $D$ waves
from Eqs.~(\ref{label25}) and (\ref{label26}) and found very good agreement
with the ones obtained from a standard partial wave solution of the deuteron
eigenvalue problem.

\section{\label{sec2}Probability Densities for Different Spin Configurations}

The operator form of the deuteron wave function given in Eq.~(\ref{label10})
is an ideal tool to express probabilities for different spin configurations
within the deuteron. This provides
analytical insight into the shape of these configurations. As an example we choose a
polarized deuteron with $ M_{d}=1 $.  Cases of interest  are if  (1)
both nucleons have their spins up, (2) both nucleons have their spins down, 
(3) one nucleon has spin up and the other has spin down, (4) one nucleon has
spin up and the other has arbitrary spin orientation and (5) one nucleon has
spin down and the other has arbitrary spin orientation. For these five cases the 
probability densities are given below. For clarity the final expressions are given
in terms of the standard $S$ and $D$ waves.

1. probability density for both nucleons having their spins up:
\begin{eqnarray}
\rho _{\uparrow \uparrow }^{1}({\bf q}) &  \equiv &  \Psi _{d}^{1\, \ast }({\bf q})\frac{1}{2}\left[ 1+\sigma _{z}(1)\right] \frac{1}{2}\left[ 1+\sigma _{z}(2)\right] \Psi _{d}^{1}({\bf q})\nonumber \\
 & = &  \frac{1}{4\pi }\left\{ \psi _{0}^{2}(q)+\frac{3}{\sqrt{2}}\left( \cos
^{2}\theta -\frac{1}{3}\right) \psi _{0}(q)\psi _{2}(q) + \frac{9}{8}\left( \cos
^{2}\theta -\frac{1}{3}\right) ^{2}\psi _{2}^{2}(q)\right\} .
\label{eq4.1}
\end{eqnarray}

2. probability density for both nucleons having their spins down:
\begin{eqnarray}
\rho _{\downarrow \downarrow }^{1}({\bf q}) &  \equiv & \Psi _{d}^{1\, \ast }({\bf q})\frac{1}{2}\left[ 1-\sigma _{z}(1)\right] \frac{1}{2}\left[ 1-\sigma _{z}(2)\right] \Psi _{d}^{1}({\bf q})\nonumber \\
 & = & \frac{9}{32\pi } \sin ^{4}\theta \: \psi _{2}^{2}(q).
\label{eq4.2}
\end{eqnarray}

3. probability density for one nucleon having spin up and the other having spin down:
\begin{eqnarray}
\rho _{\uparrow \downarrow }^{1}({\bf q})  &  \equiv & \Psi _{d}^{1\, \ast }({\bf q})\frac{1}{2}\left[ 1+\sigma _{z}(1)\right] \frac{1}{2}\left[ 1-\sigma _{z}(2)\right] \Psi _{d}^{1}({\bf q})\nonumber \\
 & = & \frac{9}{32\pi }\cos ^{2}\theta \sin ^{2} \theta \: \psi_{2}^{2}(q). \label{eq4.3}
\end{eqnarray}

4. probability density for one nucleon having spin up and the other having arbitrary spin
orientation:
\begin{eqnarray}
\rho _{\uparrow (1)}^{1}({\bf q}) &  \equiv & \Psi _{d}^{1\, \ast }({\bf q})\frac{1}{2}\left[ 1+\sigma _{z}(1)\right] \Psi _{d}^{1}({\bf q})\nonumber \\
 & = & \frac{1}{4\pi }\left[ \psi _{0}^{2}(q)+\frac{3}{\sqrt{2}}\left( \cos ^{2}\theta -\frac{1}{3}\right) \psi _{0}(q)\psi _{2}(q)\right. \nonumber \\
 &  & +\left. \frac{9}{8}\left\{ \left( \cos ^{2}\theta -\frac{1}{3}\right) ^{2}+\cos
^{2}\theta \sin ^{2}\theta  \right\} \psi _{2}^{2}(q)\right] \nonumber \\
 & = & \rho _{\uparrow \uparrow }^{1}({\bf q})+\rho _{\uparrow \downarrow
}^{1}({\bf q}). \label{eq4.4}
\end{eqnarray}

5. probability density for one nucleon having spin down and the other having arbitrary spin
orientation:
\begin{eqnarray}
\rho _{\downarrow (1)}^{1}({\bf q}) &  \equiv & \Psi _{d}^{1\, \ast }({\bf q})\frac{1}{2}\left[ 1-\sigma _{z}(1)\right] \Psi _{d}^{1}({\bf q})\nonumber \\
 & = & \frac{9}{32\pi }\sin ^{2}\theta \:  \psi _{2}^{2}(q)\nonumber \\
 & = & \rho _{\uparrow \downarrow }^{1}({\bf q})+\rho _{\downarrow \downarrow
}^{1}({\bf q}). \label{eq4.5}
\end{eqnarray}

In Figs.~\ref{fig5}-\ref{fig7} those five different probability densities are
shown.
In each figure the left side displays the probability densities as functions of
$ q $ and $ \cos \theta$,  whereas the right side depicts the probability densities
as functions of $ q_{x} $ and $ q_{z} $. The contour lines represent a vertical
section in the x-z plane through  equidensity surfaces.  Rotating this
section around the $q_z$-axis gives a three-dimensional image of the equidensity
surfaces. 

The probability densities for the first two cases, where both nucleons have the
same spin orientations, are given in Fig.~\ref{fig5}. The top row represents
$\rho _{\uparrow \uparrow }^{1}({\bf q}) $. The density  peaks at ${\bf q}=0$, 
indicating that the largest densities occur at  small momenta. This density has
a spherical shape, since Eq.~(\ref{eq4.1}) is dominated by the $S$-wave, and
in the momentum range shown has little dependence on the angle $\theta$.
The figures in the bottom row represent  $\rho _{\downarrow \downarrow }^{1}({\bf q})$. 
As Eq.~(\ref{eq4.2}) suggests, this density is only determined by the deuteron
$D$-wave times a function of the angle $\theta$. 
Thus at $\bf q$=0 it is zero, and reaches two maxima at $|q_{max}| \approx 
100$~MeV/c along the $q_x$-axis ($ \theta = \frac{\pi}{2} $).
If a measurement could be carried out on a deuteron at rest the two nucleons 
would have momenta back to back perpendicular to the polarization axis of the deuteron. 
Rotating the vertical section given in Fig.~\ref{fig5}d around the z-axis will show
a toroidal shape of the equidensity surfaces of the probability density in this configuration.
For the image in Fig.~6, two equidensity surfaces, one with a high value, being closed in the
section of Fig.~5d, and one with a small value are picked and rotated around the z-axis
resulting in a torus, being cut open vertically. The surface of lower density is left half open
at the outer side. The image displays a shape characteristic for the spherical harmonics
with $l=2, m=2$.

For the case where the spins of the two nucleons point into opposite directions,
the probability density is shown in Fig.~\ref{fig6}. According to
Eq.~(\ref{eq4.3}), this density is also given solely by the deuteron $D$-wave and a
function of the angle $\theta$. 
It has four peaks of equal
hight in each  quadrant of the $q_x-q_z$-plane at $|q_x| = |q_z| = 
q_{max} \cos (\frac{\pi}{4})$.  Rotating the vertical section in the x-z-plane
around the z-axis will reveal a double toroidal structure.
For the image in Fig.~8 two equidensity surfaces are picked and rotated around the
z-axis, resulting in a double torus being cut open vertically. The inner tubes
represent surfaces of higher density compared to the outer ones. The shape is characteristic
for a spherical harmonics with $l=2, m=1$.
Again,  a measurement  on the deuteron at rest would see in the maxima the 
two nucleons with momenta back to back pointing at $\theta=45^o$.

For the remaining two cases given by Eqs.~(\ref{eq4.4}) and (\ref{eq4.5}), 
where only one of the two nucleons is polarized, 
 the probability densities are presented in Fig.~\ref{fig7}. 
The figures in the top row represent $\rho _{\uparrow (1)}^{1}({\bf q})$. 
For the momentum range shown its  properties are very similar to 
$\rho _{\uparrow \uparrow}^{1}({\bf q})$ given in the top row of Fig.~\ref{fig5}. The
reason is that $\rho _{\uparrow \uparrow}^{1}({\bf q})$ is 
larger than $\rho _{\uparrow \downarrow}^{1}({\bf q})$ and thus dominates. 
The figures in the bottom row depict
$\rho _{\downarrow (1)}^{1}({\bf q})$. This density  has the same maxima as 
$\rho _{\downarrow \downarrow}^{1}({\bf q})$ given in Fig.~\ref{fig5}d, 
but a slightly different angular behavior. 
For a fixed $q$  the changes with $ \theta $ are slower than for 
$\rho _{\downarrow \downarrow}^{1}({\bf q})$. This is  caused by the linear 
dependence on 
$\sin^2 \theta $, 
whereas $\rho _{\downarrow \downarrow}^{1}({\bf q})$ has a quadratic one.
A rendering of two equidensity surfaces is displayed in Fig.~10, which shows the
different angular behavior when comparing to Fig.~6.

\section{Summary}

As an object with internal structure it is tempting to investigate the deuteron properties 
three dimensionally. To that aim we study the deuteron properties in a representation
based on the total helicity of the two-body system taken along the relative momentum of
the two particles. 
Though originally developed for describing NN scattering, the method is 
general and can be used to solve bound state problems as well. 

We introduce deuteron wave function components in the helicity basis. 
They depend on the magnitude $q$ of the relative momentum and the angle $\theta$ 
of the relative momentum ${\bf q}$ to the z axis. Deriving an `operator form' of 
the deuteron wave function 
one obtains insight into the analytical angular behavior of those components.

We derived two sets of two coupled eigenvalue equations for deuteron wave functions. 
The first set of equations does not use  any a priori knowledge of the quantum numbers
of the deuteron, and the NN potential representation in helicity basis is used similarly 
as for NN scattering. As a consequence one has coupled two dimensional equations. 
In the second set of equations,  the well known deuteron properties of a $S$ and $D$ 
waves content is build in and the analytical angular behavior of those amplitudes is 
taken into account explicitly. In this case, one arrives at one dimensional equations.
This second derivation bears some similarity to partial wave methods. The two one-dimensional 
amplitude components are each linear combination of the standard $S$ and $D$ wave function components.

The calculated deuteron binding energy determined in both ways agrees perfectly 
with the value determined in standard calculations based on partial wave representations
of the deuteron eigenvalue equation. The newly defined helicity wave function components 
depend on Wigner's d-function and the $q$ dependent part are linear combinations of 
the standard $S$ and $D$ waves of the deuteron. 
We display their properties for different projections of the total angular momentum 
$M_d$. Like for NN scattering we can connect the helicity amplitudes to the 
standard $S$ and $D$ waves  and find perfect numerical agreement with the partial 
wave components determined in a standard manner. 

Finally we evaluate various spin and momentum dependent probabilities in a fashion,
 which is exact with respect to the angular dependence. This is made possible by using 
the  `operator form' of the deuteron state. 
It is conceivable that in quasi elastic electrodesintegration of the deuteron 
one may be able to see those momentum dependent spin distributions. 

Summarizing,  we extended a recently introduced helicity representation for 
NN scattering to the NN bound state.  This formulation leads to new forms for deuteron 
wave function components, which can be determined by two coupled equations.

\vfill

\acknowledgements This work was performed in part under the
auspices of the Deutsche Akademische Austauschdienst under contract No. A/96/32258,
the U.~S.  Department of Energy under contract
No. DE-FG02-93ER40756 with Ohio University, the NATO Collaborative
Research Grant 960892, and the
National Science Foundation under Grant No. INT-9726624.
We thank the Neumann Institute for Computing (NIC) at the Forschungzentrum J\"ulich and the
Computer Center of the RWTH  Aachen (Grant P039) for the use of their
facilities. We especially thank H. Schumacher from NIC for his assistance with the
three-dimensional images.

\noindent
\begin{figure}
\caption{\label{fig1}The deuteron wave function components $ \varphi ^{0}_{0}(q,\theta )$
(a) and $ \varphi ^{0}_{1}(q,\theta )$ (b) in units $ 10^{-3}$ MeV$ ^{-1.5}$ as functions
of $ q$ and $ \cos \theta $.}
\end{figure}

\noindent
\begin{figure}
\caption{\label{fig2}The deuteron wave function components $ \varphi ^{1}_{0}(q,\theta )$
(a), $ \varphi ^{1}_{1}(q,\theta )$ (b), $ \varphi ^{-1}_{0}(q,\theta )$ (c) and 
$ \varphi ^{-1}_{1}(q,\theta )$ (d) in units $ 10^{-3}$ MeV$ ^{-1.5}$
as functions of $ q$ and $ \cos \theta $.
The relation between the different components is discussed in Section~\ref{sec4}.}
\end{figure}

\noindent
\begin{figure}
\caption{\label{fig3}The deuteron density for $ M_{d}=0$ (a and b)
and $ M_{d}=1$ (c and d) in units $ 10^{-6}$ MeV$ ^{-3}$ as functions of $ q$ and
$ \cos \theta $ (a and c) and as functions of $ q_{x}$
and $ q_{z}$ (b and d). For the momentum range shown the deuteron densities display
a uniform angular behavior. The contours represent equidensity lines along a
vertical section in the x-z plane.}
\end{figure}

\noindent
\begin{figure}
\caption{\label{fig4} The wave functions $|\Phi _{0}(q)|$ and $ |\Phi _{1}(q)|$
in units MeV$ ^{-1.5}$.}
\end{figure}

\noindent
\begin{figure}
\caption{\label{fig5}The probability densities $ \rho _{\uparrow \uparrow }^{1}({\bf q})$ in units $ 10^{-6}$
MeV$ ^{-3}$ for both nucleons having their spins up
 (a and b) and $ \rho _{\downarrow \downarrow }^{1}({\bf q})$ in units
$ 10^{-10}$ MeV$ ^{-3}$ for both nucleons having their spins down (c and d).
The contours represent equidensity lines along a
vertical section in the x-z plane.}
\end{figure}

\noindent
\begin{figure}
\caption{Two selected equidensity surfaces of the probability density 
$ \rho _{\downarrow \downarrow}^{1}({\bf q})$ for both nucleons having their spins down. 
The image is created by rotating two of the equidensity lines of 
Fig.~5~d around the z-axis. Note that the z-axis is  stretched with respect to the
other two axes. }
\end{figure}

\noindent
\begin{figure}
\caption{\label{fig6}The probability density $ \rho _{\uparrow \downarrow }^{1}({\bf q})$
in units $ 10^{-10}$ MeV$ ^{-3}$ for one nucleon having spin up and the other having 
spin down. The contours represent equidensity lines along a
vertical section in the x-z plane.}
\end{figure}

\noindent
\begin{figure}
\caption{Two selected equidensity surfaces of the probability density 
$ \rho _{\uparrow \downarrow }^{1}({\bf q})$ for one nucleon having spin up and 
the other having spin down.
The image is created by rotating two of the equidensity lines of Fig.~7~b around the 
z-axis.}
\end{figure}

\noindent
\begin{figure}
\caption{\label{fig7}The probability densities $ \rho _{\uparrow (1)}^{1}({\bf q})$
in units $ 10^{-6}$ MeV$ ^{-3}$ for one nucleon having spin up whereas 
the other having arbitrary spin orientation (a and b) 
and $ \rho _{\downarrow (1)}^{1}({\bf q})$ in units $ 10^{-10}$
MeV$ ^{-3}$ for one nucleon having spin down whereas the other having arbitrary spin 
orientation (c and d). The contours represent equidensity lines along a
vertical section in the x-z plane.}
\end{figure}


\begin{references}
\bibitem{1}I. Fachruddin, Ch. Elster, and W. Gl\"ockle, Phys. Rev. C \textbf{62}, 044002
(2000).

\bibitem{Kubis} M. Jacob, and G.C. Wick, Ann. Phys. (N.Y.), 404 (1959).

\bibitem{alzetta} R. Alzetta, K. Erkelenz, K. Holinde, Nucl. Phys. \textbf{A185}, 459 (1972).

\bibitem{bonnb}R. Machleidt, Adv. Nucl. Phys. \textbf{19}, 189 (1989).

\bibitem{4}J. L. Forest,  V.R. Pandharipande, S.C. Pieper, R.B. Wiringa, R.
Schiavilla, A. Arriaga,  Phys. Rev. C \textbf{54}, 646 (1996).

\bibitem{av18} R. B. Wiringa, V. G. J. Stoks, R. Schiavilla, Phys. Rev. {\bf C51}, 38 (1995).

\bibitem{ericson} T. Ericson and W. Weise, `Pions and Nuclei', The
International Series of Monographs in Physics {\bf 74}, Claredon Press, Oxford
1998, p. 49. 

\bibitem{5}P. L. DeVries, \textit{A First Course in Computational Physics} (Wiley, New York,
1994).
\bibitem{6}R. A. Malfliet and J. A. Tjon, Nucl. Phys. \textbf{A127}, 161 (1969).
\bibitem{7}W. Gl\"ockle, Nucl. Phys. \textbf{A381}, 343 (1982).

\bibitem{9} M. E. Rose, \textit{Elementary Theory of Angular Momentum} (Wiley, New York,
1957).
\end{references}
\end{document}